\documentstyle[12pt,epsf]{article}
\begin{document}
\def\dfrac#1#2{{\displaystyle {#1 \over #2}}}
\renewcommand{\thefootnote}{\fnsymbol{footnote}}
\textwidth 5.5in
\textheight 8.5in
\pagestyle{empty}
\parskip.5cm

\begin{center}
{\rm \Large New Physics Effects on Higgs Production \\ 
at $\gamma \gamma $ Colliders\footnote{Work 
supported by CONACyT.}\footnote{To appear in the 
Proceeding of the {\it V Mexican 
Workshop of Particles and Fields}, Puebla, M\'exico, October 1995.
}}

\vskip.5cm
{\rm \large J. M. Hern\'andez$^a$, M. A. P\'erez$^a$ and J. J. Toscano$^b$}
\end{center}

{\it $a$. Departamento de F\'\i sica, CINVESTAV-IPN, Apdo. Postal 14-740, 
07000, M\'exico, D. F., M\'exico.}
 
{\it $b$. Facultad de Ciencias F\'\i sico Matem\'aticas, Universidad 
Aut\'onoma de Puebla, Apdo. Postal 1152, 72000, Puebla, Pue., M\'exico.}
 
\bigskip
\medskip
 
Abstract. We study heavy physics effects on the Higgs production in 
$\gamma \gamma $ fusion using the effective Lagrangian approach. 
We find that the effects coming from
new physics may enhance the standard model predictions for the number of
events expected in the final states $\bar bb$, $WW$, and $ZZ$ up to one
order of magnitude, whereas the corresponding number of events for the final
state $\bar tt$ may be enhanced up to two orders of magnitude. 
\vskip15pt

In the search of the nature and source of the electroweak symmetry 
brea\-king a decisive point will be the set in operation of the LHC and the 
next generation of linear $e^+ e^-$ collider (NLC). The NLC will 
permit us to use the old idea of Compton laser backscattering in order to 
reach a center of mass energy of $\sqrt{s}=200-500$ GeV [1] in the
$\gamma \gamma $ mode.

In particular, the $\gamma \gamma $ colliders offer
a great opportunity to study the dynamics of the elusive Higgs
boson. If the standard model (SM) Higgs is detected  through its dominant 
decay modes in $e^+ e^-$ or $p \bar p$ collisions, then a $\gamma \gamma $ 
collider will allow a direct measurement of its partial decay width 
into two photons. The $H^{\circ }\gamma \gamma $ interaction is an 
one-loop prediction of the SM and all their extensions, in which all 
the contents of
charged particles participates. Thus, a precise measurement of its decay 
width will permit us to
discriminate between the SM and new physics predictions.

In the present work we study possible effects of new physics on the Higgs 
boson production in $\gamma \gamma $ collisions within the context of the
effective Lagrangian approach. The framework of effective 
Lagrangians, as a mean to parametrize physics beyond
 the SM in a model-independent manner, have been extensively discussed 
in the recent literature in both
decoupling and nondecoupling cases [2]. In this work we consider the 
decoupling case [3], where the SM can be obtained as a low-energy limit 
of a weakly coupled renormalizable full theory. 
The heavy fields effects are parametrized by a series of
high-dimensional nonrenormalizable operators, constructed out of the SM
fields. These operators respect the SM symmetries [4], and because that,
 it is possible to establish the 
order of perturbative theory in which they may be generated in the full 
theory [5]. 

Since for $m_{H^{\circ }}\leq 400$ GeV the total width of the Higgs boson is
small in comparison with the energy of the photons beam, the number of 
$H^{\circ }\rightarrow X$ expected events may be written as: 
\begin{equation}
\label{events}N=\left[ \dfrac{dL_{\gamma \gamma }}{dW_{\gamma \gamma }}
\right] _{m_H}\dfrac{4\pi ^2}{m_H^2}\Gamma \left( H^{\circ }\rightarrow
\gamma \gamma \right) BR\left( H^{\circ }\rightarrow X\right) \left(
1+\lambda \tilde \lambda \right) ,
\end{equation}
where $BR\left( H^{\circ }\rightarrow X\right) $ is the branching ratio of
the Higgs boson into the final state $X$ ( $X=\bar bb$, $WW$, $ZZ$, $\bar 
tt$ ), 
$\lambda $ and $\tilde \lambda $ are the helicity states of the scattered
photons, $(dL_{\gamma \gamma }/dW_{\gamma \gamma })$ is the differential 
$\gamma \gamma $ luminosity as a function of the two-photon invariant mass.
Since the $H^{\circ }\gamma \gamma$
interaction is generated at one-loop level by the dimension-four theory, a
complete calculation requires to consider all the nonrenormalizable
operators which contribute at this level in the full theory. 
All these contributions to the decay $H^{\circ
}\rightarrow \gamma \gamma $, including the SM contribution, are 
suppressed by
the loop factor $(16\pi ^2)^{-1}$. Consequently, if the new physics scale 
$\Lambda $ is not very far from the Fermi scale $v=\sqrt{2}\langle \phi \rangle
_{\circ }$, the contributions of dimension eight operators, which may be
generated at tree-level by the underlying physics, are dominant with respect
to the contributions of the operators of dimension six [5-6]. 
Accordingly, 
the relevant effective Lagrangian for the Higgs boson production in $\gamma
\gamma $ fusion may be written as: 
\begin{equation}
\label{Lagrangian}
\begin{array}{c}
{\cal L}_{\rm{eff}}={\cal L}_0+\dfrac 1{\Lambda ^2}\left[ \alpha _{b\phi
}O_{b\phi }+\alpha _{t\phi }O_{t\phi }+\alpha _\phi ^{(1)}O_\phi
^{(1)}+\alpha _\phi ^{(3)}O_\phi ^{(3)}\right] \\ + \dfrac 1{\Lambda
^4}\left[ \alpha _{8,1}O_{8,1}+\alpha _{8,3}O_{8,3}+\alpha
_{8,5}O_{8,5}\right] ,
\end{array}
\end{equation}
where ${\cal L}_0{\cal \ }$is the SM Lagrangian and $\alpha _i$ are
unknown parameters; the operators of dimension eight and dimension six
 which induce the $H^{\circ }\gamma \gamma $ and $H^{\circ }q {\bar q}$ 
vertices at tree-level, are given in [6-7].
Using the Lagrangian (1) we obtain the following expressions for the partial
decay widths,

\begin{equation}
\label{width1}\Gamma _{\rm{eff}}\left( H^{\circ }\rightarrow \gamma 
\gamma \right) =\dfrac{\alpha ^2G_Fm_H^3}{64\sqrt{2}\pi ^3}\left|
\sum_fQ_f^2N_cF_0^f+F_0^b+\left( \dfrac v\Lambda \right) ^4F_{a8}^b\right|
^2, 
\end{equation}
\begin{equation}
\label{width2}\Gamma _{\rm{eff}}\left( H^{\circ }\rightarrow 
W^{+}W^{-}\right)
=\left[ 1+\dfrac 14\left( \dfrac v\Lambda \right) ^2\left( 2\alpha _\phi
^{(1)}-\alpha _\phi ^{(3)}\right) \right] ^2\Gamma_{SM}\left( H^{\circ
}\rightarrow W^{+}W^{-}\right) , 
\end{equation}
\begin{equation}
\label{width3}\Gamma _{\rm{eff}}\left( H^{\circ }\rightarrow ZZ\right) 
=\left[ 1+\dfrac 12\left( \dfrac v\Lambda \right) ^2\left( \alpha _\phi
^{(1)}+\alpha _\phi ^{(3)}\right) \right] ^2\Gamma_{SM}\left( H^{\circ
}\rightarrow ZZ\right) , 
\end{equation}
\begin{equation}
\label{width4}\Gamma _{\rm{eff}}\left( H^{\circ }\rightarrow \bar 
qq\right)
=\left[ 1-\dfrac{\sqrt{2}m_Z}{C_0m_q}\left( \dfrac v\Lambda \right) ^2\alpha
_{q\phi }\right] ^2\Gamma_{SM}\left( H^{\circ }\rightarrow \bar qq\right) , 
\end{equation}

\noindent where $q$ stands for the $b$ and $t$ quarks, $Q_f$ is the electric 
charge of the fermions, $N_c$ is 3 for quarks and 1 for leptons, 
$C_0=m_Z\sqrt{\sqrt{2}G_F}$. 
The parametric functions $F_0^f$ and $F_0^b$ 
associated to fermions and $W$ boson loops, respectively,
are given in Ref.[8], $F_{a8}^b$ is given in [7].

In Fig. 1-2 we display the number of expected events as a function of the
Higgs boson mass for the final states $\bar bb$ and $\bar tt$. 
The corresponding figures 
for the final states $WW$ and $ZZ$ are shown in [7], they show the 
same behaviour as the $\bar bb$ case. We
have taken all the coefficients of the nonrenormalizable
interactions as $\mid \alpha _i\mid =1$ and  the new physics scale 
as $\Lambda =1$ TeV.
We consider a center of mass energy of the $e^{+}e^{-}$ collider
equal to 500 GeV.The luminosity of the $e^{+}e^{-}$ beam is taken
equal to 20fb$^{-1}$. Futhermore, we have consi\-de\-red the theoretical
limit $\lambda \tilde \lambda =1$.

We present only
the two extreme scenarious corresponding to a maximum and a minimum
number of events for the total contribution, which includes effective
interactions in all the vertices involved. We have
checked that for $\Lambda \geq 3$ TeV, the heavy physics effects
decouples from the SM predictions. In the scenario with a
maximum number of events we can appreciate an enhancenment of almost one
order of magnitude, with respect to the SM predictions, for the 
final state $\bar bb$, $WW$ and $ZZ$, while for the $\bar tt$ channel 
the enhancenment may be up to two orders of magnitude [7]. In 
the less favorable
scenario the $\bar bb$ final state is suppressed by one 
order of magnitude
with respect to the SM prediction, while in the $\bar tt$
final state subsist a remarkable enhancenment with 
respect to the SM
predictions.

\begin{center} 
\leavevmode 
\epsfysize=8cm \epsfbox{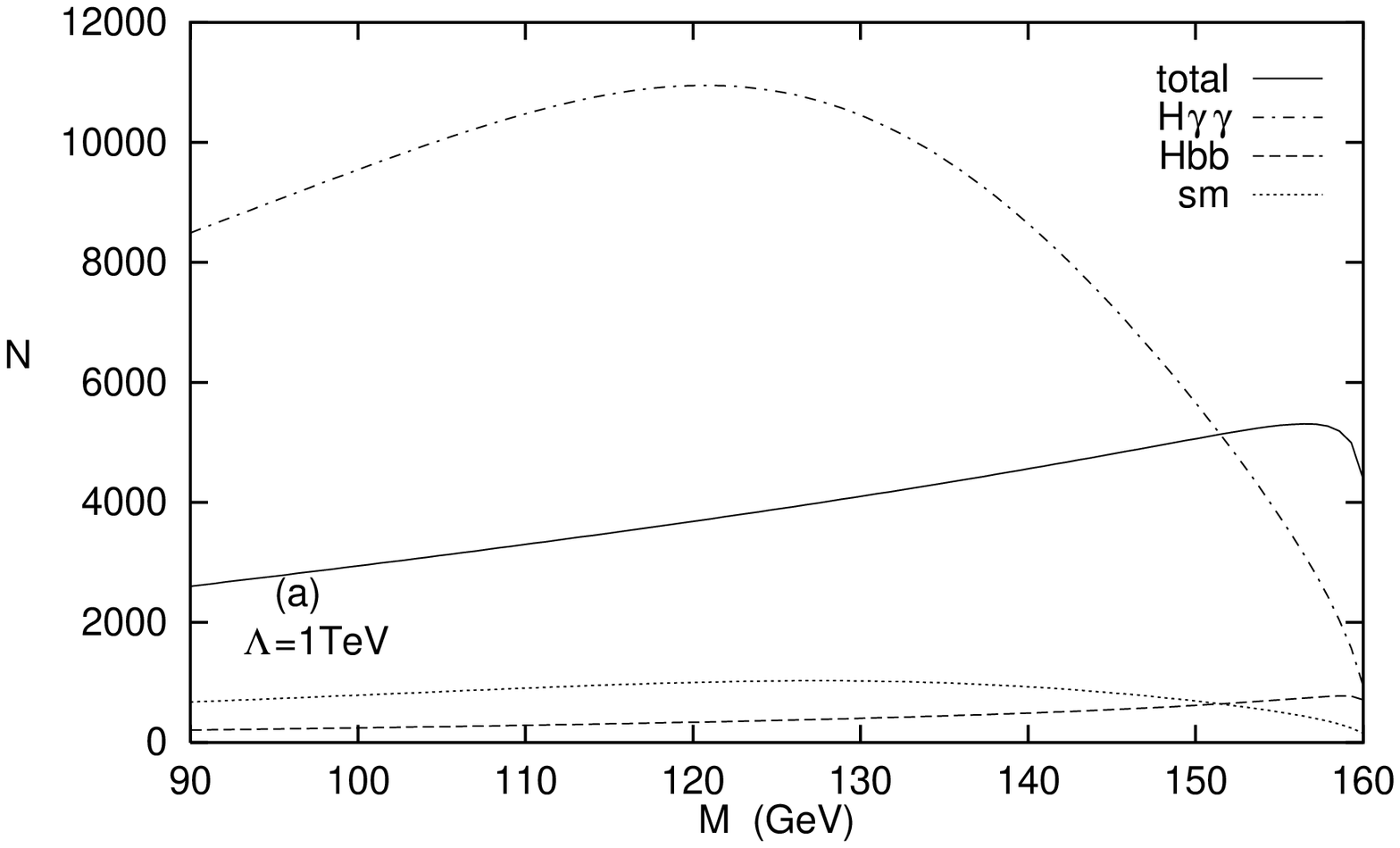}

\leavevmode 
\epsfysize=8cm \epsfbox{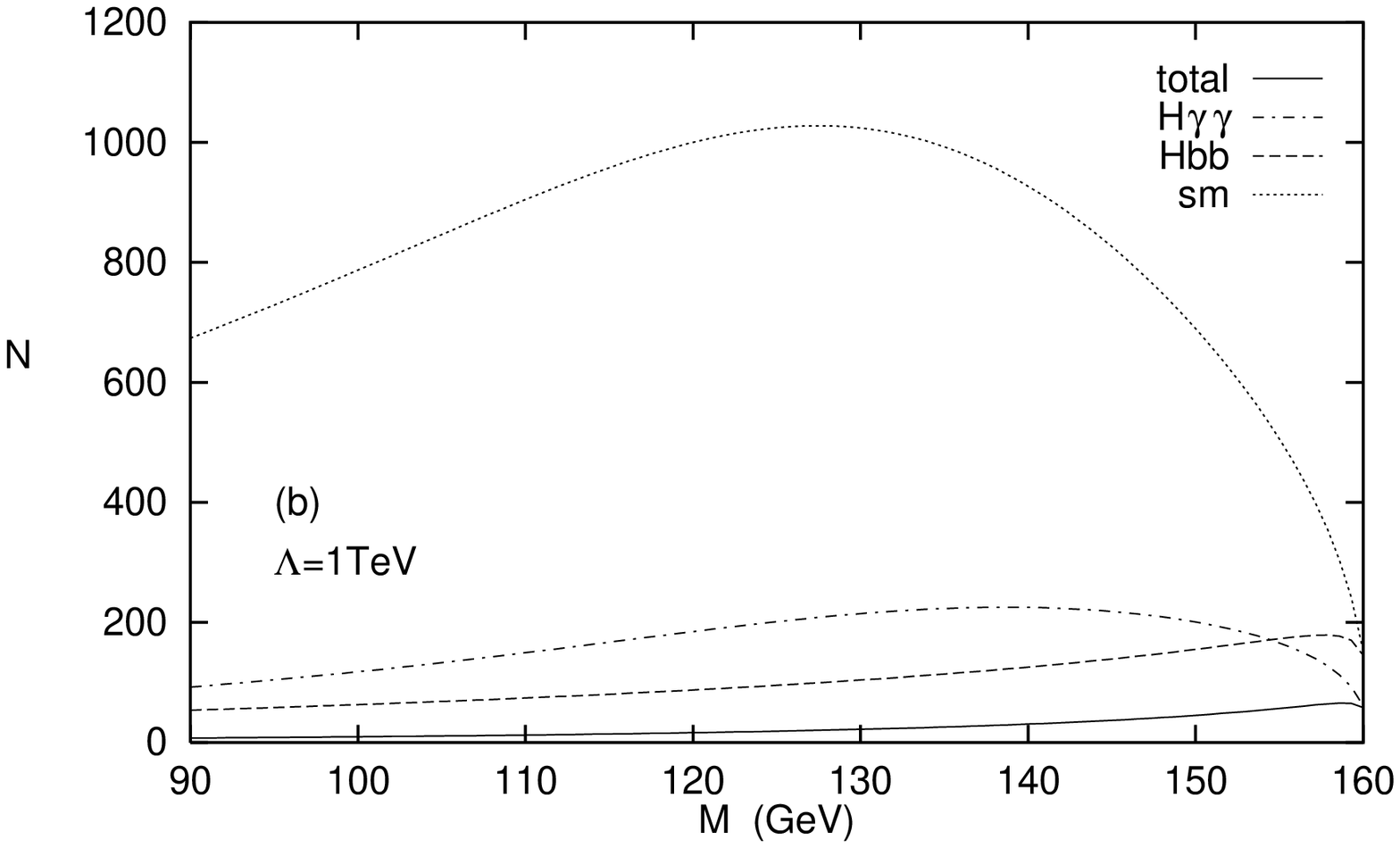}
\end{center}
 
Fig.1. Number of events expected for the $\bar bb$ final state as a
function of the Higgs boson mass and $\Lambda =1$ TeV, for the scenarious
with a maximum (a) and minimum (b) number of events.

\begin{center}
\leavevmode
\epsfysize=8cm \epsfbox{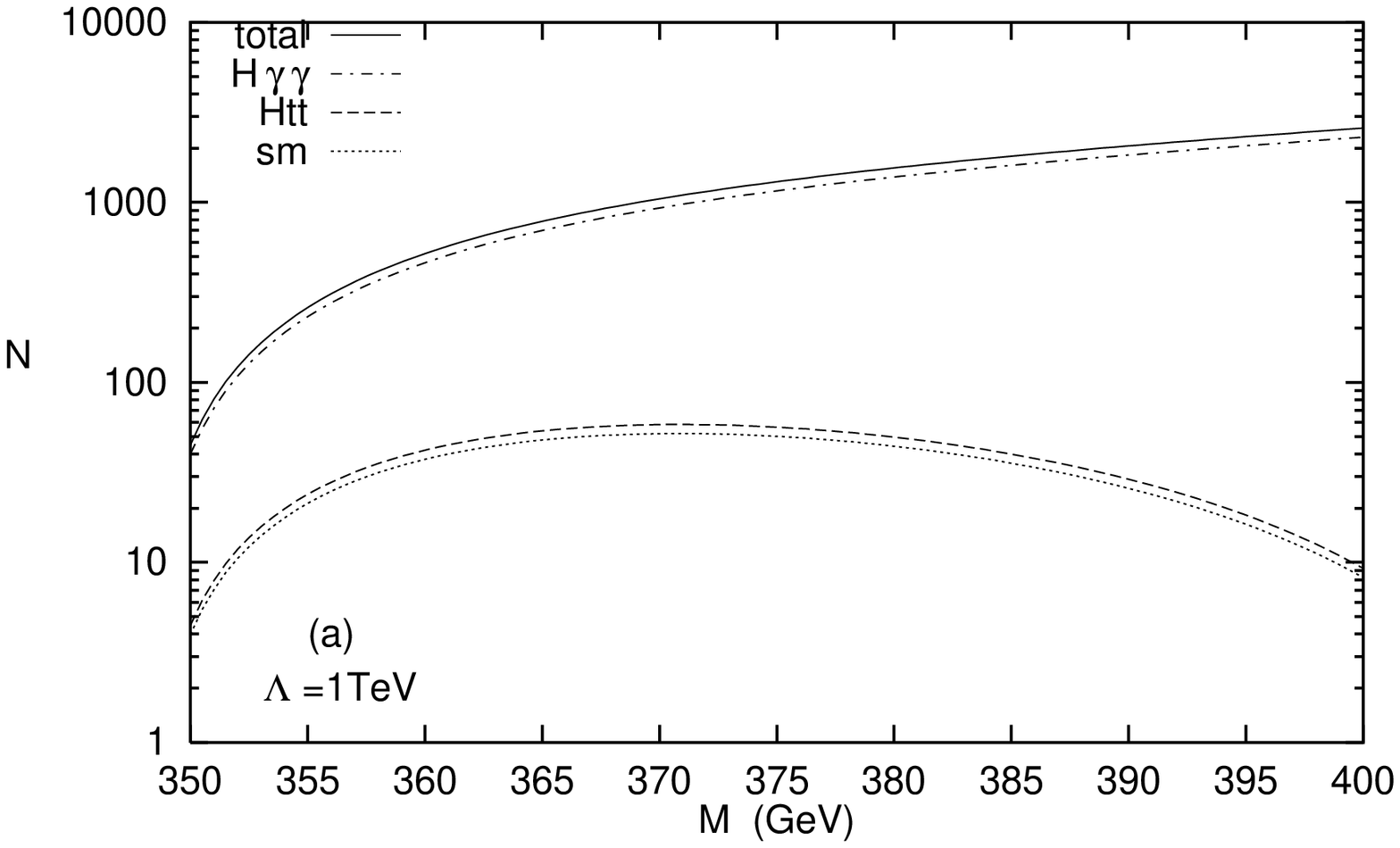}
 
\leavevmode
\epsfysize=8cm \epsfbox{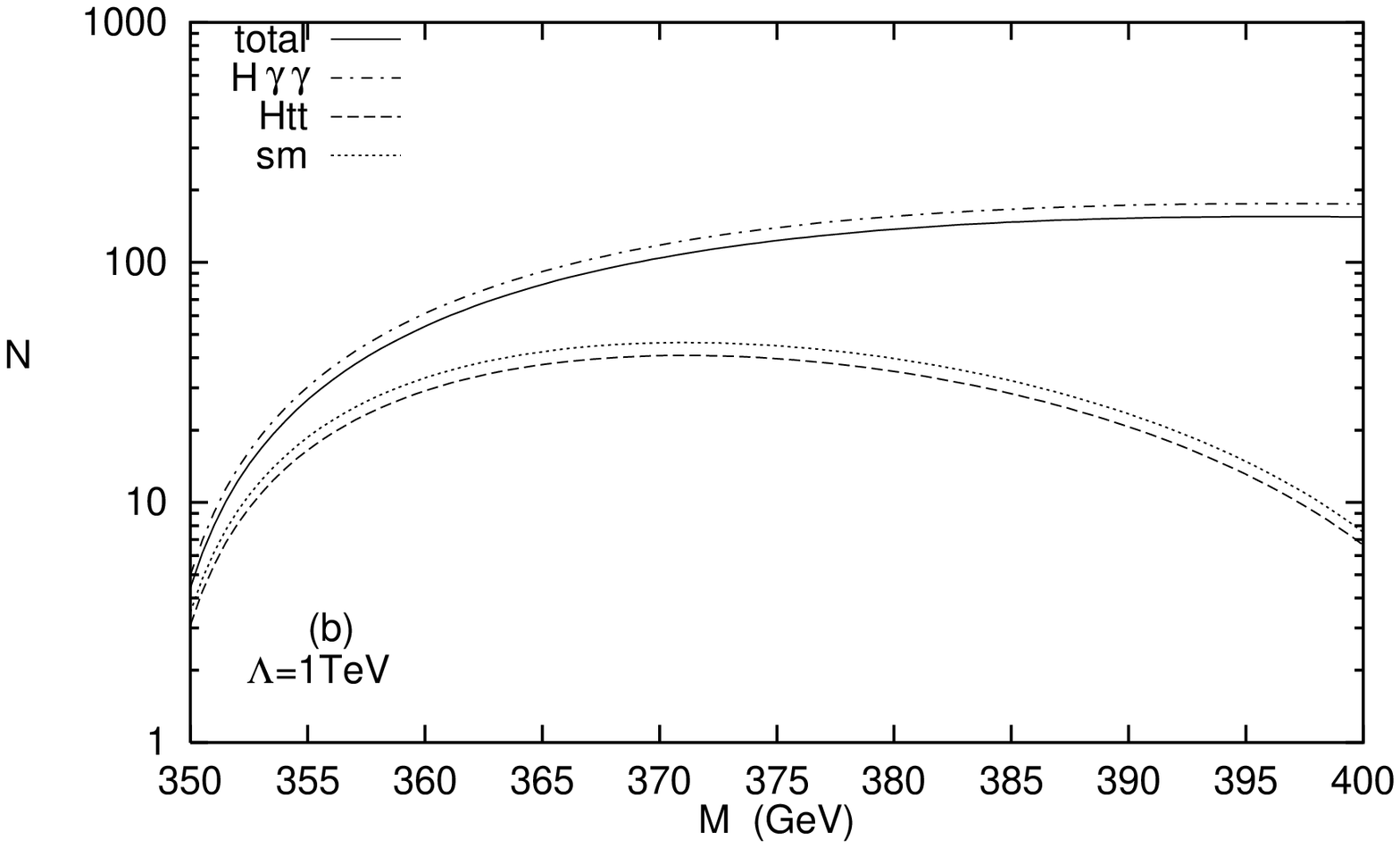}
\end{center}
 
Fig. 2. Number of events expected for the $\bar tt$ final state as a
function of the Higgs boson mass and $\Lambda =1$ TeV, for the scenarious
with a maximum (a) and minimum (b) number of events.
\newpage

In conclusion, the Higgs boson production through $\gamma \gamma $ fusion
results in a highly sensitive mechanism to detect physics beyond the SM. 
Futhermore, we expect that the enhancement induced by  new physics on the 
signal is not hidden by the corresponding effects on the 
background.\footnote{This issue is discussed more extensively in [7].}

\end{document}